\documentclass[12pt,preprint]{aastex}

\newcommand{\Ion}[2]{#1{\,\footnotesize #2}}

\shorttitle{{\it{FUSE}} Spectroscopy of VW Hyi in Quiescence}
\shortauthors{Godon et al.}

\begin{document}

\title{Far Ultraviolet Observations of the Dwarf Nova VW Hyi in
Quiescence 
\footnote{Based on observations made with the NASA-CNES-CSA {\it{Far
Ultraviolet Spectroscopic Explorer}}. {\it{FUSE}} is operated for NASA
by the
Johns Hopkins University under NASA contract NAS5-32985}}  

\author{Patrick Godon\altaffilmark{2}, Edward M. Sion}
\affil{Astronomy and Astrophysics, Villanova University, \\ 
800 Lancaster Avenue, Villanova, PA 19085, USA}
\email{patrick.godon@villanova.edu edward.sion@villanova.edu}

\author{F. H. Cheng}
\affil{Department of Physics, Shanghai University, 
99 Shang-Da Road, Shanghai 200436,
P.R. China}
\email{cheng3@prodigy.net}

\author{Paula Szkody}
\affil{Department of Astronomy,
University of Washington,
Seattle, WA 98195, USA}
\email{szkody@alicar.astro.washington.edu}

\author{Knox S. Long}  
\affil{Space Telescope Science Institute, 3700 San Martin Drive,
Baltimore, MD 21218, USA}
\email{long@stsci.edu} 

\author{Cynthia S. Froning}
\affil{Center for Astrophysics and Space Astronomy, 
University of Colorado, 593 UCB, Boulder, CO 80309, USA} 
\email{cfroning@casa.colorado.edu}

\altaffiltext{2}
{Visiting at the Space Telescope Science Institute, Baltimore, MD 21218,
USA}

\clearpage 

\begin{abstract}

We present a 904-1183 \AA\ spectrum of the dwarf nova VW Hydri
taken with the {\it{Far Ultraviolet Spectroscopic Explorer}} 
during quiescence, eleven days after a normal outburst, when the
underlying white dwarf accreter is clearly exposed in the far 
ultraviolet. 
However, model fitting show that a uniform temperature white dwarf
does not reproduce the overall spectrum, especially at 
the shortest wavelengths.
A better approximation to the spectrum is obtained with a model
consisting of a white dwarf and a rapidly rotating ``accretion belt''. 
The white dwarf component accounts for 83\% of
the total flux, has a temperature of $23,000$K, a $v \sin{i} = 400$ 
km~s$^{-1}$, and a low carbon abundance.  
The best-fit accretion belt component accounts 
for 17\% of the total flux, 
has a temperature of about $48,000-50,000$K, 
and a rotation rate $V_{rot} \sin{i}$ around $3,000-4,000$ km~s$^{-1}$.  
The requirement of two components in the modeling of the
spectrum of VW Hyi in quiescence helps to resolve some of the
differences in interpretation of ultraviolet
spectra of VW Hyi in quiescence.
However, the physical existence of a second component
(and its exact nature) in VW Hyi itself
is still relatively uncertain,
given the lack of better models for spectra of the inner disk in
a quiescent dwarf nova.

\end{abstract}


\section{Introduction: Accretion Onto White Dwarfs in Dwarf Novae} 

\subsection{Dwarf Novae} 

Dwarf novae (DNe) are a subclass of cataclysmic variable (CV) systems,
in which a white dwarf (WD, the primary) accretes hydrogen-rich matter
from a low-mass main sequence-like star (the secondary) filling its
Roche lobe.
In these systems, the transferred gas forms an accretion disk around the WD.  
It is believed that the accretion disk is subject
to a thermal instability that causes cyclic changes of the accretion
rate.
A low rate of accretion ($\approx 10^{-11} M_{\odot}$~yr$^{-1}$)
quiescent stage is followed every few weeks to
months by a high rate of accretion ($\approx 10^{-8} M_{\odot}$~yr$^{-1}$)
outburst stage of days to weeks.
These outbursts (dwarf nova - DN accretion event or nova-like high
state),
are believed to be punctuated every few thousand years or more by a
thermonuclear runaway (TNR) explosion: the classical nova \citep{hac93}. 

The WD dominates the far ultraviolet (FUV) in many, and probably, most,
DNe in quiescence. 
As a consequence, quiescent DNe provide a unique laboratory 
not only for understanding the physics of  
accretion but also for understanding physical processes of
WDs. As a result, 
{\it{Hubble Space Telescope}}  ({\it{HST}}) and other 
UV observatories have been used to  directly observe
the effects of accretion on the WDs of these systems. These
studies have yielded determinations of effective temperature,
the rotation rates, photospheric abundances, cooling rates
following outburst, and dynamical masses in a number of DNe
\citep{sio99,gan99}.

However, there are some indications of an additional component
besides the white dwarf in the spectra of some DNe in quiescence,
such as the presence of emission lines and the bottoms of Lyman 
alpha profiles which do not go to zero as in a pure white dwarf. 
Possible locations of this additional component could be: 
(1) a heated region of the WD; 
(2) direct emission from the boundary layer; 
(3) an optically thick region of the disk; or 
(4) a corona/chromosphere above a cool disk 
\citep{kok89,kok92,mey89,ko96,liu97,lad97}.  

\subsection{VW Hyi} 
VW Hyi is a key system for understanding DNe in general. 
It is the closest \citep[placed it at 65 pc]{war87}
and brightest example of an SU UMa-type DN and it lies
along a line of sight with an exceptionally low interstellar column
\citep[estimated the HI column to be 
$\approx 6 \times 10^{17}$ cm$^{-2}$]{pol90},
which has permitted study of VW Hyi in nearly all wavelength ranges,
including detection in the usual opaque extreme ultraviolet
[EUVE \citep{mau96}].  
For these reasons, it is one of the best-studied systems. 
Coherent and quasi-coherent soft X-ray oscillations and a 
surprisingly low luminosity boundary layer (BL) have been detected
\citep{bel91,mau91}.  
VW Hyi is below the CV period gap near its lower edge,
with an orbital period of 107 minutes and a quiescent 
optical magnitude of 13.8. 
It is a member of the SU UMa class of DNe, which undergo 
both normal DN outbursts and superoutbursts. The normal 
outbursts last 1-3 days and occur every 20-30 days,
with  peak visual magnitude of 9.5. The superoutbursts 
last 10-15 days and occur every 5-6 months, with peak 
visual magnitude reaching 8.5. The mass of the accreting 
WD was estimated to be 0.63 $M_{\odot}$ \citep{sch81}, 
but more recently a gravitational redshift determination 
yielded a larger mass $M_{wd}=0.86 M_{\odot}$ \citep{sio97}. 
Below the period gap, gravitational wave emission is 
thought to drive mass transfer, resulting in very low accretion
rates during dwarf nova quiescence.  
The inclination of the system is $\approx 60$ degrees \citep{hua96a,hua96b}. 

VW Hyi was first observed at FUV wavelengths with {\it IUE}.  Based on
observations of VW Hyi in quiescence, which showed a broad absorption
profile centered on Lyman $\alpha$, \citet{mat84} argued that
FUV light from the system was dominated by the WD with 
$T_{eff} = 18,000 \pm 2,000$ K 
(for log g=8). Much higher S/N spectra were obtained
with {\it HST}.  In particular, \citet{sio95,sio96,sio01} confirmed
the basic shape of the spectrum and concluded that the temperature of
the WD varied by at least 2,000 K depending on the time since outburst.
They
also presented evidence for CNO processing of material in the WD
photosphere from the relatively narrow metal lines in the spectrum, as
well as for a rapidly rotating ``accretion belt'' in the system. The
accretion belt was to be understood physically as a region of the WD
surface spun up by accreting material with Keplerian velocities.
Furthermore, \citet{sio97}, using the GHRS on {\it HST}, measured
the gravitational redshift of the WD and concluded that the mass of the
WD was $0.86 ^{+0.18}_{-0.32} M_{\odot}$ and that the rotation rate of
the WD
was $\sim$ 400 km~s$^{-1}$.  All of these observation were limited to a
wavelength range longward of about 1150 \AA. However, an 820-1840 \AA\
spectrum of VW Hyi was obtained using the Hopkins Ultraviolet Telescope
(HUT).
\citet{lon96} found that the HUT spectrum was reasonably
consistent with a WD with a temperature of about 17,000K, but that an
improved fit to the data at that time could be obtained with a
combination of emission from a WD and an accretion disk.

\citet{lon96} suggested that higher S/N spectra of VW Hyi in quiescence
are ``needed to unambiguously assess the disk contribution to the FUV
spectrum of VW Hyi in quiescence''. Here we have attempted to follow-up
on that suggestion by observing VW Hyi in quiescence with the {\it Far
Ultraviolet Spectroscopic Explorer} ({\it{FUSE}}).  
With a practical wavelength range
of 904-1188 \AA, FUSE is sensitive to a second component, since the
expected flux from a WD with a temperature of about 20,000 K is very
different at 950 and 1100 \AA.  In addition, the {\it FUSE} spectral
range and high spectral resolution allows the study of a broad range of
line transitions.  

The remainder of the paper is organized as follows.  In
Section 2, we describe the observations and provide a description of the
spectra that were obtained.  In Section 3, we compare the spectrum to
models in an attempt to decompose the spectrum into its constituent
parts.  In Section 4,  we discuss the origin of the second component in
the spectrum, and in Section 5, we briefly summarize our conclusions.

\section{Observations and Data Reduction} 

\subsection{The Observations} 

{\it{FUSE}} is a low-earth orbit satellite, launched in June 1999
\citep{moo00}. Its optical
system consists of four optical telescopes (mirrors), each separately 
connected to a different Rowland spectrograph. The 4 diffraction
gratings
of the 4 Rowland spectrographs produce 4 independent spectra on two
photon 
counting area detectors. The total spectral wavelength coverage extends
from 904 \AA\ to 1188 \AA\ with a resolution of $R\approx 12000$.
Two mirrors and two gratings are coated
with SiC to provide wavelength coverage below 1020 \AA, while the other
two mirrors and gratings are coated with Al and a LiF overcoat.  The
Al+Lif coating provides about twice the reflectivity of SiC at 
wavelengths $ > 1050$ \AA, and very little reflectivity below 1020 \AA\
(hereafter the SiC1, SiC2, LiF1 and LiF2 channels).
The observatory, its operations, and performance have been described
in detail by \citet{sah00}.  

The {\it{FUSE}} observation of VW Hyi occurred on August 29th, 2001
at 16:48 UT (JD2452151) approximately 11 days after the last outburst.
The
overall exposure time was 18,004 seconds through the LWRS aperture,
acquired in 9 exposures ranging from $\simeq$600 to 3500 seconds each.  
The data were obtained in time-tag mode. Of the 9
exposures acquired, all but the first two experienced a series of event
bursts \citep{sah00} that were strongest in
segments LiF1A and LiF1B but affected all eight segment spectra.  We
re-processed the data using V.2.1.6 of the CALFUSE pipeline, which
includes
screening and removal of event bursts during data reduction. We combined
the resulting exposures and data segments to create a time-averaged
spectrum
with a linear, 0.1 \AA\ dispersion, 
weighting the flux in each output datum by
the exposure time and sensitivity of the input exposure and channel of
origin.  The S/N is about 10:1 at a resolution of 0.1 \AA. Due to the
``worm'',  the data in the LiF1 channel longward of $\approx 1130$ \AA\
were removed. Because of that, the practical spectral range of
{\it{FUSE}} was reduced to 905-1182 \AA.

\subsection{The Spectrum} 
In Figure 1, the {\it{FUSE}} spectrum is displayed in a flux versus
wavelength
plot with the principal line features labeled with tick marks. 
An interesting aspect
of the spectrum is that while the region
longward of 1010 \AA\ shows features that can clearly be identified as
absorption lines, the region shortward of this does not.
The most obvious absorption lines are NII (1085.7 \AA ), 
SiIII (at 1113.2, 1144.3, and 1160.2 \AA ) and the (familiar)
CIII (1174.9-1176.4 \AA ).  
In table 1, the line measurements (equivalent widths, full widths at
half-maximum, ion identifications and wavelengths) are presented.
In addition, 
there is an emission feature redward of Lyman $\beta$, at 1033 
\AA\ and 1037 \AA, that we identify as OVI emission lines.   

We note that the flux level of the {\it{FUSE}} spectrum, in the range of
wavelength overlap with STIS E140M spectra (i.e. around 
1150-1182 \AA ) , matches rather closely (within about 10\%) 
the HST flux value of $\approx 2 \times 10^{-13}$ erg~s$^{-1}$cm$^{-2}$\AA$^{-1}$. 
And a comparison of the {\it{FUSE}} spectrum to that obtained with HUT 
\citep{lon96} indicates   
(a) that the overall spectral shape
appears rather similar, but (b) that the FUSE spectrum clearly shows
emission extending to the Lyman limit.  
There is clearly a rise in the flux in
the region of the Lyman limit. 

\subsection{Variability} 
We also examined the variability among the 9 exposures in the
observations by calculating the standard deviation about the 
mean flux in three wavelength regions: 
917 -- 936 \AA, 953 -- 970 \AA, and 1050 -- 1070 \AA.  
These three regions sample the spectrum at the short wavelength
emission peak, slightly longward of the emission peak, and to 
the red of Lyman beta. In a single channel
(SiC2), the standard deviation about the mean for the 9 exposures was
21\%,
15\%, and 8\%, respectively, indicating that the FUV spectrum of VW Hyi
does fluctuate, with the variability strongest at the shortest
wavelengths,
where the WD does not contribute.

\section{Analysis}

In this section we describe the procedure we follow to assess  
the temperature, rotation rate and chemical abundances of the
accreting WD in VW Hyi. 
This procedure consists of comparing the
observed FUSE spectrum of VW Hyi with a grid of theoretical 
spectra obtained
assuming different assumptions and for different values of the
parameters of the system as explained below. 
For this purpose, we use a combination of synthetic
stellar, disk and accretion belt spectra. 
The best fit models
are then obtained using a $\chi^{2}$ minimization fitting
procedure. 
   
The stellar atmosphere models and the accretion belt models 
are both generated using 
the TLUSTY code \citep{hub88} for different values of the parameters 
of the accreting WD, such as temperature, composition and 
surface gravity. A spectrum synthesis code (SYNSPEC) 
is then used to generate 
the spectra of the stellar atmospheres obtained in TLUSTY 
\citep{hub94,hub95}. 

For the accretion disk spectrum, we used the grid of accretion disk spectra 
computed by \citet{wad98}, who use a slightly different version of 
the TLUSTY code (TLUSDISK) to generate the theoretical spectrum of 
an accretion disk. The accretion disk model is really made of a 
collection of rings. The disk models are computed assuming LTE and 
vertical hydrostatic equilibrium. Irradiation from external sources
is neglected. Local spectra of disk annuli are computed taking into 
account line transition from elements 1-28 (H through Ni). Limb 
darkening as well as Doppler broadening and blending of lines are 
taken into account. 

Then, to carry out the model fits, we masked the following
wavelength regions where several narrow
emission-like features occur in the {\it{FUSE}} spectrum:  
[948-952\AA], 
[972-974\AA], 
[989-993\AA], 
[1025-1041\AA], 
[1078-1082\AA], 
[1092-1096\AA], 
[1168-1170\AA].
Most of these emission lines are due to air glow, when part of the 
observations are carried out during day time. In particular we masked
the OVI emission lines around 1033 \AA\ and 1037 \AA\ .  

Details of the codes and the $\chi^{2}_{\nu}$ 
($\chi^2$ per degree of freedom) 
minimization fitting procedures 
are discussed in detail in \citet{sio95} and \citet{hua96a}, 
and will not be repeated here.

\subsection{The white dwarf model}

We took the WD photospheric temperature T$_{eff}$, log g, Si and
C abundances, and rotational velocity V$_{rot}\sin{i}$ as free parameters. 
Since our WD model spectra are normalized to 
1 solar radius at a distance of 1 kiloparsec, the distance $d$ of a
source is related to the scale factor $S$:  
$$d = 1000(pc) \times (R_{wd}/R_{\sun})/\sqrt{S},$$ and the scale
factor $S$ is the factor by which the theoretical fluxes 
have to be multiplied in 
order to fit the observed fluxes. In the first model presented here
we fixed $d=65$ pc while the radius (and consequently the mass) of the WD 
was allowed to change. However, other models presented here have been computed
assuming a fixed radius (corresponding to $M_{wd}=0.86 M_{\odot}$).  

The grid of models for the
one component (white dwarf only) extended
over the following range of parameters: 
T$_{wd}/1000$K= 22, 23, ....., 36; 
Si = 0.1, 0.2, 0.5, 1.0, 2.0, 5.0 (times solar); 
C = 0.1, 0.2, 0.5, 1.0, 2.0, 5.0 (times solar); 
V$_{rot}\sin{i} $= 100, 200, 400, 600, 800 km~s$^{-1}$; and
$\log{g}=7.5$ to $9.0$ by increment of $0.1$.

For the single temperature white dwarf models, 
we found the best-fitting results with 3-sigma
error bars to be $T_{wd}/1000$(K)  = 26.0 + 0.2/-0.4, 
Si abundance = 0.5  + 0.3/-0.1 $\times$ solar,
C abundance = 0.1  + 0.2/-0.1 $\times$ solar, 
$\log{g}=8.8$ and 
$V_{rot}\sin{i}= 400$+ 100/-200 km~s$^{-1}$. This best-fit
has a $\chi^2_{\nu}$ = 14.81 with a mass of
$M_{wd}=1.14 M_{\odot}$. The best 
single-temperature white dwarf model fit is displayed
in figure 2 (see also Table 2: model 1).  

Following the suggestion of an anonymous referee, 
we also modified our fitting code to 
keep both the distance (65 pc) and the radius 
($7 \times 10^8$ cm for a $0.86 M_{\odot}$, or
equivalently $\log{g}=8.37$) fixed at the same time. 
These parameters defined a single value of 
the scale factor $S$.
In this way, we proceeded to find the best fitting 
models by varying the temperature and 
composition while keeping log g fixed. 

In Table 2 we present 4 WD models (models \# 2, 3, 4, \& 5) 
with a fixed $\log{g}=8.37$ and a 
fixed distance of 65 pc. The best model has a $T=22,000$K, and a
$\chi^2_{\nu}=28.2$. Here we have assumed solar abundances. Changing
the abundances modestly improves the $\chi^2_{\nu}$ value.  

Thus an interesting comparison became possible 
with our original fitting procedure where the scale factor 
was variable. In the case of the WD only, the fixed scale factor
models do not improve the fit, and the best fit model has a 
$\chi^2_{\nu}$ value twice as large as the one obtained when  
allowing $S$ to vary.  
We used the same masked regions 
as we did in the variable scale factor fitting.
We should note, however, that while we assumed $M=0.86 M_{\odot}$,
the mass estimate from \citet{sio97} of $0.86 M_{\odot}$ has an error
bar of $^{+0.18}_{-0.32} M_{\odot}$. This makes the mass
and the radius of the WD known within a factor of 2:
$0.54 < M_{wd}/M_{\odot} < 1.04$ with a WD radius corresponding to 
6,000km$< R_{wd} <$12,000km,
or equivalently $7.5 < \log{g} < 8.56$, which justifies our
original fitting procedure where we kept $\log{g}$ and consequently
$M_{wd}$ and $R_{wd}$ as free parameters. 

\subsection{The accretion disk model}

Due to the relatively poor fit using single temperature WD models,
particularly the flux deficit of the model at short wavelengths, 
we explored whether an accretion disk model would produce better
agreement with the {\it{FUSE}} spectrum. 
Although one expects the inner region of the accretion disk in VW Hyi 
to be optically thin during quiescence, the lack of such models led 
us to consider how well optically thick steady state disks can 
represent the {\it{FUSE}} observation.

In the present work we used the grid of accretion disk spectra
of \citet{wad98}  
consisting of 26 different combinations 
of $M_{wd}$ (0.35, 0.55, 0.80, 1.03 and 1.21 $M_{\odot}$) and 
$\dot{M}$ ($log \dot{M}$ = -8.0, -8.5, -9.0, -9.5, -10.0 and -10.5 
$\dot{M}$ yr$^{-1}$; see Table 2 in \citet{wad98}) and the spectra 
are presented for six different disk inclinations $i$ (8.1, 18.2, 41.4, 
60.0, 75.5 and 81.4 degrees),
in the form of nonprojected flux (however, the projection of the Keplerian 
velocities is included) scaled to a distance of 100 pc. 
Here, we corrected for the projected flux and the distance
is related to scale factor: $$ d=100(pc)/\sqrt{S}.$$   

First, we fixed only the distance d=65 pc and kept $M_{wd}$ as a 
free parameter.  
The best-fitting accretion disk model (model \# 6 in Table 2)
corresponded to $M_{wd}=1.03 M_{\odot}$, 
$\dot{M} = 1 \times 10^{-10.5} M_{\odot}$/yr$^{-1}$,  
inclination angle i = 81 degrees.
With a $\chi^2_{\nu}$ value of 19.9, 
this disk only fit is significantly worse than the 
single-temperature white dwarf fit (see figure 3). 
This should not be surprising 
since the white dwarf is the dominant FUV emitter in the system 
\citep{mat84,sio95,sio96, sio97,sio01,lon96}. 

Next, we fixed the distance d=65 pc and chose $M=0.8 M_{\odot}$
(in better agreement with the mass estimate of $0.86M_{\odot}$).  
The best-fitting accretion disk model (model \# 8 in Table 2)
corresponded to $\dot{M} = 1 \times 10^{-10} M_{\odot}$/yr$^{-1}$,  
inclination angle i = 60 degrees with a $\chi^2_{\nu}$ value of 
29.83. In Table 2 we present 2 additional models (7 \& 9 in Table 2), 
one with a higher accretion rate 
and one with a lower accretion rate, to emphasize how drastically
the $\chi^2_{\nu}$ value can worsen from one model to another
when changing the accretion rate.  

\subsection{Composite model: WD plus accretion disk} 

Next we tried a combination of a WD plus an optically thick disk.
We created a grid in WD temperature $T_{eff}$ from
16,000K to 35,000K in steps of 1000K, with each time a mass $M_{wd}$
in agreement with a white dwarf mass the same as each one assumed in the
grid of disk models of
\citet{wad98}. 
The disk flux is divided by 100 to normalize it at 
1000pc to match the WD flux, then both fluxes are added for comparison with
the observed flux. In that cases one has: 
$$ F_{obs} = \left[ F_{wd} \left( R_{wd}/ R_{\odot} \right)^2 +
F_{disk}/100 \right] \times S; $$
where $F_{obs}$ is the observed FUSE flux (integrated over the entire
FUSE wavelength range), $F_{wd}$ is the theoretical (integrated) flux
of the WD, $F_{disk}$ is the theoretical (integrated) flux of the
disk,  and $d=1000/\sqrt{S}$ pc. 

Here too, we started by fixing $d=65$ pc.   
The best-fitting disk plus WD combination, 
yielded $T_{eff} =
23,000K$, V$_{rot}\sin{i} 
= 400$km~s$^{-1}$, Si = 0.5 x solar, C = 0.1 x solar,
$M_{wd}=1.21 M_{\odot}$,  with a disk model having an  
accretion rate $10^{-10.5} M_{\odot}$yr$^{-1}$ and $i=81$ degrees. 
The combination fit had a $\chi^2_{\nu}$ = 11.03. 
In this fit, the WD contributed 64\% of the flux and the
accretion disk 36\% of the flux. This composite model is shown
in Figure 4 and is listed in Table 2 (model 10).  

We then tried a model in which we fixed both d=65pc and the radius
$R_{wd}$ of the WD. The disk models (from the grid of models of \citet{wad98})
with parameters in best agreement with
the parameters of the VW Hyi systems are the disk models with $M_{wd}=
0.8 M_{\odot}$ (corresponding to $\log{g}=8.23$) and $i=60$ degrees. 
The best such composite WD plus disk fit model is presented in
Table 2 (model 11), with $T_{wd}=21,000$K, 
$\dot{M}=10^{-10.5} M_{\odot}$yr$^{-1}$,
$V_{rot}\sin{i}=400$km~s$^{-1}$ with a $\chi^2_{\nu}=32.9$. Here too
we assumed solar abundances for simplicity.  
The WD contributes 88\% of the flux while the disk contributes 12\%
of the flux. 
 
\subsection{Composite model: WD plus accretion belt} 

Lastly, we tried a combination of a WD plus an accretion belt. 
The accretion belt model is really a stellar atmosphere model
but with only a fractional area of the WD and parameters consistent
with a fast rotating hot accretion belt. Since the belt is fast rotating
its effective gravity is reduced to a fraction of the gravity of the
star as follows:  
$$ g_{belt} = g_{wd} - \frac{V_{belt}^2}{R_{wd}}.$$ 

Once $\log{g_{wd}}$ is set,  
there is a direct relation between the gravity of the 
belt $\log{g_{belt}}$ and its velocity $V_{belt}$. 
For the accretion belt plus WD 
composite models, we created a grid in WD temperature $T_{eff}$ from
16,000K to 35,000K in steps of 1000K, and in accretion belt temperature
$T_{belt}$ from 25,000K to 55,000K in steps 
of 1000K.
However, the results
are not very sensitive to the value of $\log{g_{belt}}$ 
as long as $\log{g_{belt}} < 7.5$, corresponding to 
$V_{belt} \sin{i} \approx 3,000-4,000$ km~s$^{-1}$,
where the lower limit corresponds to $\log{g_{wd}}=8.37$ (a $0.86 M_{\odot}$
WD) and the upper limit corresponds to $\log{g_{wd}}=8.54$ (a
$0.96 M_{wd}$ WD), and we have assumed $i=60$ degrees.  

In the case of the WD plus belt the distance $d$ is given in a way similar to 
the WD only case with 
$$ d=1000(pc) \times (R_{wd}/R_{\odot}) / \sqrt{S},$$
where we set d=65 pc.  

The best-fitting white dwarf plus 
accretion belt fit with fixed distance (d=65pc) and a
free radius $R_{wd}$  
yielded $T_{wd} = 23,000$K, Si abundance = 2.0 $\times$ solar, 
C abundance = 0.2 $\times$ solar, $V_{rot}\sin{i}= 400$km~s$^{-1}$,
with a radius $R_{wd}=0.0087R_{\odot}$, corresponding to a mass 
$M_{wd}=0.96 M_{\odot}$ or $\log{g}=8.54$.  
The belt temperature was $T_{belt} = 48,000$K, with a
velocity of $V_{belt}\sin{i} \approx 4,000$km~s$^{-1}$.  
In this model, the white dwarf contributed 83\% of 
the FUV flux and the accretion belt 17\% of the FUV flux, 
and the fractional area of the accretion belt was 2\%. The 
$\chi^2_{\nu}$ value of this fit was 7.06.
This best-fitting white dwarf plus accretion 
belt composite model is displayed in figure 5 and 
in Table 2 (model 12).
We find this composite fit to be clearly 
superior to the single temperature and accretion disk
- only fits. The remarkable lowering of the reduced 
$\chi^2_{\nu}$ value for the accretion belt fit
is additional confirmation of the findings 
by \citet{sio96,sio97,sio01} and \citet{gan96}  
that VW Hyi's 
white dwarf has an inhomogeneous temperature 
distribution and that the most likely explanation 
is that of an accretion belt of higher 
temperature at its equatorial latitudes. 

On the other hand, the best-fitting combination WD plus accretion belt model 
with a fixed WD radius and distance (model 13 in Table 2, $M=0.86 M_{\odot}$),
had a $\chi^2_{\nu}$ = 10.9. The
stellar and belt parameters corresponding to this best-fit are as
follows.
The WD has an average surface temperature $T_{eff}$ = 22,000K, 
$V_{rot}\sin{i}= 400$km~s$^{-1}$, 
with solar abundances. The accretion belt has the
parameters
$T_{belt}$ = 50,000K, and $V_{belt} \sin{i}= 3,000$km~s$^{-1}$. 
The belt area is 2\% of the WD surface and the 
belt contributes 23\% of the total flux while the WD 
contributes 77\% of flux.
This composite two-temperature fit is shown in figure 6. 

It is interesting to note that irrespective of whether
the 
radius (and consequently the mass) of the white dwarf was kept
fixed or not in the models, the best fit models with a lower 
$\chi^2_{\nu}$ and with parameters ($M_{wd}$, $d$, $i$) 
consistent with the values assessed for the system 
were obtained for the WD plus belt composite model (models 12
\& 13). 
The reason the fit is better is that the belt, with its high
temperature,
accounts for the flux shortward of 950 \AA (where the WD does not
contribute),
and its high rotational velocity matches better the rather featureless 
spectrum of the second component. Model 10 (WD plus disk) 
has a $\chi^2_{\nu} \approx 11$,
however its WD mass is too large $M_{wd}=1.21$ and the
inclination is too high at i = 81 degrees, therefore it can
be ruled out. However, its relatively low
$\chi^2_{\nu}$ might be due to the fact that the hot inner disk is the main
component in the FUV with (Table 2 in \citet{wad98}) a rotational
velocity $V_{rot} \sin{i}\approx 5,000$km~s$^{-1}$ and a peak temperature
$T_{max}\approx 30,000$K.

\section{Discussion and Conclusion}

The main observational 
result in this work is the confirmation that the {\it{FUSE}} 
spectrum of VW Hyi in quiescence cannot be modeled as a single
WD temperature, and that at least two components are needed
in the modeling of the data.  
The dominant component is that of a $23,000$K WD, while
the second component is some kind of true featureless
continuum with a color temperature that is higher than that of the WD. 
It is possible that 
the two components exist only in the modeling of the FUV data and not
in VW Hyi itself, for example if the WD is not heated homogeneously 
then we would have one component with a continuous range of temperatures
that contribute to the spectrum.  

\subsection{The Cooling of the WD} 
VW Hyi has been previously observed mainly during quiescence, following 
normal (3 days) outburst and superoutburst (about 2 weeks). From an IUE 
archival study, \citet{gan96} assessed that the WD cools down
exponentially to a mean temperature $T_{wd} \approx 19,000K$, 
with an exponential 
decay time of $\approx 3$ days after normal outburst and $\approx 10$
days
after superoutburst. From this archival data alone it appears that
the maximum temperature the WD reaches is about 23,000K,   
just after the end of a normal outburst and about 26,000K just after
the end of a superoutburst. In Table 3 we recapitulate the recent
observations of VW Hyi with different instruments. Most of the
temperature
estimates for the accreting WD are in the range described in
\citet{gan96}.

In the present work, the best fitting model to the 
{\it{FUSE}} data is a combination of a WD plus an accretion belt, 
which is noticeably better than the WD alone or the disk models. 
These two components to the spectrum of VW Hyi in quiescence 
help to resolve some of the differences in interpretation 
of UV spectra of VW Hyi in quiescence.
The first component has 
a photospheric temperature of 23,000K, a rotation
rate of 400 km~s$^{-1}$ and chemical abundances that 
are reasonably consistent with previous HST
FOS, GHRS and STIS results. This seems to indicate a cooling of the
WD 11 days after a normal 3 day outburst. 
The second component has an effective temperature higher than that
of the WD, and a featureless (rather flat) spectrum.   

\subsection{The Nature of the Second Component}

Our numerical modeling is unable to account 
in detail for some of the features such as:    
(1) large discrepancies between our theoretical spectrum and the
present FUSE data around 925 \AA\ and 1100 \AA;  
(2) the accretion belt model does not account for the upturn in
flux at the 
Lyman limit, where the variability is strongest, which is also the
region of the spectrum where the WD does not contribute to the flux, 
and consequently, neither the WD nor the accretion belt model 
can possibly be the source of the variability; 
(3) the  OVI emission lines in the {\it{FUSE}} 
spectrum indicate the possible presence of an 
optically thin source in the system.  

Therefore, the exact nature of the second component 
is still relatively uncertain, due to  
the lack of better 
models for spectra of the inner disk in a quiescent dwarf nova. 
As we mentioned in \S 1  the possible candidates for a 
second component are:\\  
(1) an accretion belt, i.e. a fast rotating  
heated layer of the surface of the WD,
possibly created in the outburst and which remains hot primarily 
because of the effective viscosity of the WD. \\ 
(2) the boundary layer, i.e. the region between the inner
edge of the disk and the stellar surface where the remaining
kinetic energy of the accreting flow is released. This optically
thin region reflects the instantaneous accretion rate and is expected
to emit X-ray. \\ 
(3) an optically thick disk, however physically unjustified in the
accretion disk limit cycle model \citep{can93},
is what we are numerically able to model. \\ 
(4) or a corona/chromosphere above a cool disk. \\ 
At the present time, only the optically
thick accretion belt and accretion disk can be numerically 
modeled, while no detailed modeling exists for the optically thin 
components (spectra). The results we obtained here indicate 
that the fast rotating belt is the best candidate for the 
source of the second FUV component.  

\subsection{The Accretion Belt} 

While the accretion belt model has improved the fits in VW Hyi in many
observations \citep{sio01}, its existence has not yet been established.  
At some point, it was suggested [see e.g. \citep{hua96a}] that the second
hot component might actually be a hot ring near the inner edge of the
disk. \citet{hua96a} identified the hot ring component as the ``hot
state'' suggested in the accretion disk limit cycle model \citep{can93}.
In that theory, the outburst is triggered near the inner edge of the disk,
and a heating front propagates to the outer edge, transforming the entire
disk to the hot state. The quiescent state is achieved when a cooling
front propagates back to smaller radii and shuts off the flow onto the WD.
In that particular case, one expects the characteristics of the second
component (observed in quiescence) to change rapidly in time, as the front
moves inward through the disk. Recent observations \citep{sio01}, however,
show that the second component in HST/STIS spectra of VW Hyi, taken 5 days
apart, remains pretty much the same, making the scenario of the ring
unlikely. 

While accretion belts were first discussed theoretically by
\citep{kip78,kut87,kut89}, the first observational detection was for U
Gem's WD during quiescence \citep{lon93} based on {\it{HUT}} observations
followed VW Hyi's WD by \citet{sio96} using HST and \citet{gan96}
using IUE archival spectra. The physical basis for an
accretion belt is the tangential accretion of disk matter at the stellar
equator with spin-up of the surface layers of the WD as it shears into the
WD envelope with the slow conversion of kinetic energy to heat as a result
of viscous heating in the differentially rotating atmosphere.

We have confirmed that the modeling of the {\it{FUSE}} 
spectrum of VW Hyi in quiescence requires at least two
components, as had been suggested by the earlier HST and
HUT observations. The main component in the modeling is identified as
the 
photosphere of a WD with a  temperature of about $23,000$K.  
The second component is a continuum relatively featureless with an
effective
temperature $\approx 48,000$K.

This discussion mainly serves to emphasize how far we are from
an understanding of VW Hyi without a knowledge of the location
of the emitting regions within the system. There could be a continuous
range of temperatures, along with ranges in emitting areas and 
velocities, that contribute to the observed FUV spectrum. 
The only firm result that has been obtained here is that the 
second modeling component is fairly flat, and one possibility to produce
such a flat component is with a high rotational velocity and a high
temperature. 

\acknowledgements 
This work was supported by a Cycle 2 NASA {\it{FUSE}} 
grant to Villanova University.

\clearpage

{\bf{Figures Caption}}
 
Figure 1: {\it{FUSE}} spectrum of VW Hyi with lines identified.      
The sharp emission lines
are not intrinsic to the source, they are due to air glow.

Figure 2: 
{\it{FUSE}} spectrum of VW Hyi together with the synthetic spectrum for
a single white dwarf model with T=26,000K, $V_{rot}\sin{i}=400$km~s$^{-1}$, 
and the composition as specified in the text (model 1 in Table 2). 

Figure 3:
{\it{FUSE}} spectrum of VW Hyi together with the synthetic spectrum for
an accretion disk model with a $1.03M_{\odot}$ central star, 
$\dot{M}=1.0 \times 10^{-10.5} M_{\odot}$yr$^{-1}$, and a disk inclination of
81 degrees (model 6 in Table 2).

Figure 4:
{\it{FUSE}} spectrum of VW Hyi together with the synthetic spectrum for
a composite model fit consisting of a WD (dotted line) and an 
accretion disk (dashed line). 
The WD model has $T_{eff} = 23,000$K, $V_{rot}\sin{i}=400$km~s$^{-1}$ 
and the
disk has an accretion rate of $3.16 \times 10^{-11} 
M_{\odot}$yr$^{-1}$ and 
an inclination of 81 degrees. 
The combination fit is shown with the solid line.
(model 10 in Table 2)

Figure 5: 
{\it{FUSE}} spectrum of VW Hyi together with the synthetic spectrum 
for a composite  model fit consisting of a WD and an accretion belt.  
The white dwarf model has $M_{wd}=0.96 M_{\odot}$,  
T=23,000K, $V_{rot}\sin{i}=400$km~s$^{-1}$,
and the composition as specified in the text, together with an
accretion belt with $T=48,000$K, and $V_{belt}\sin{i}=4,000$km~s$^{-1}$.
The white dwarf photosphere flux is shown with the dotted line,
the accretion belt flux is shown with the dashed line
and their combined flux is shown with the solid line.
(model 12 in Table 2)

Figure 6: 
{\it{FUSE}} spectrum of VW Hyi together with the synthetic spectrum 
for a composite  model fit consisting of a WD and an accretion belt.  
The white dwarf model has $M_{wd}=0.86 M_{\odot}$,  
T=22,000K, $V_{rot}\sin{i}=400$km~s$^{-1}$,
and solar composition, together with an
accretion belt with $T=50,000$K, and $V_{belt}\sin{i}=3,000$km~s$^{-1}$.
The white dwarf photosphere flux is shown with the dotted line,
the accretion belt flux is shown with the dashed line
and their combined flux is shown with the solid line.
(model 13 in Table 2)

\clearpage

\begin{figure}
\plotone{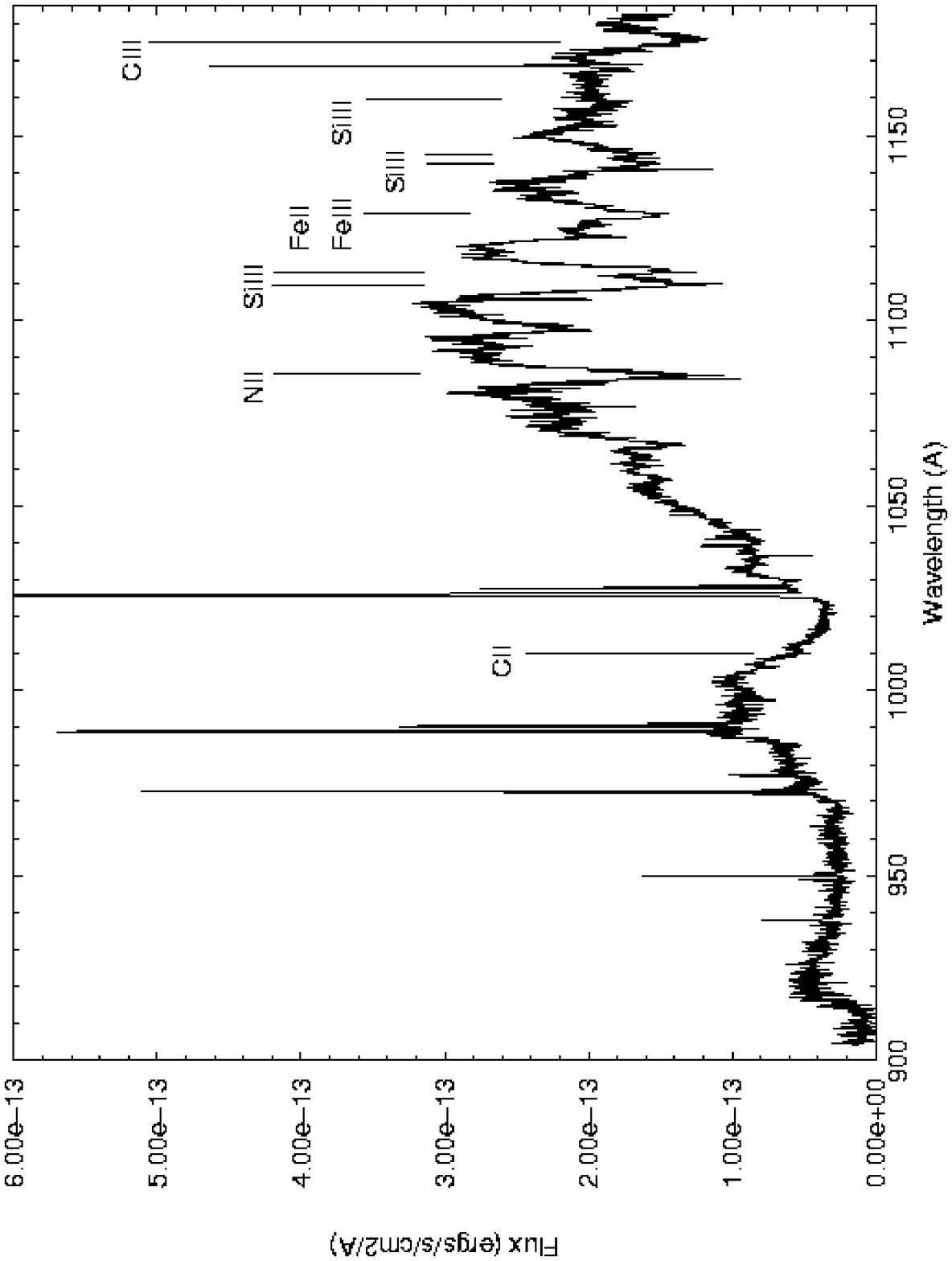}
\figurenum{1}
\end{figure}

\clearpage

\begin{figure}
\plotone{f2.eps}
\figurenum{2}
\end{figure}

\clearpage
\newpage 

\begin{figure}
\plotone{f3.eps}
\figurenum{3}
\end{figure}

\clearpage
\newpage 

\begin{figure}
\plotone{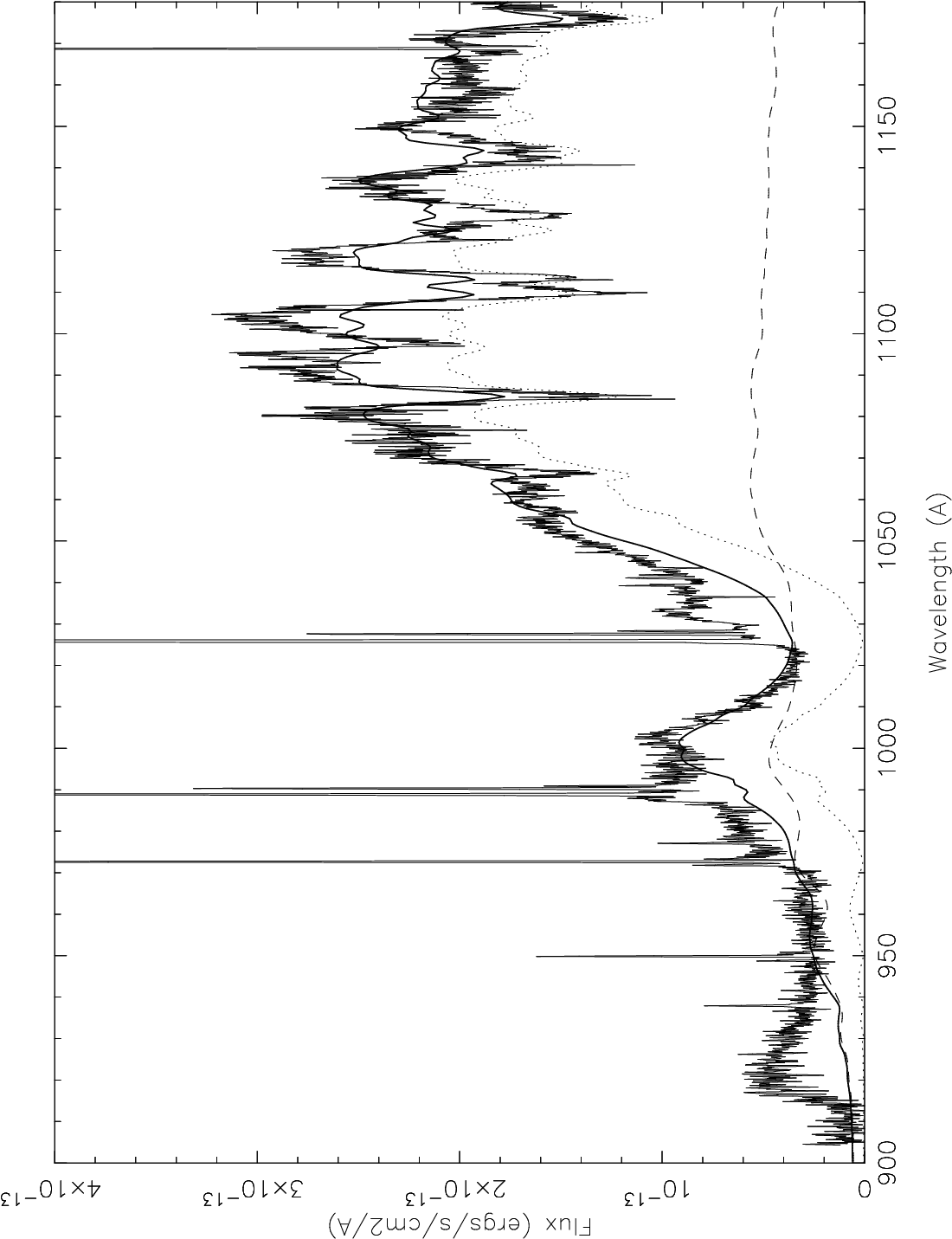}
\figurenum{4}
\end{figure}

\clearpage
\newpage 

\begin{figure}
\plotone{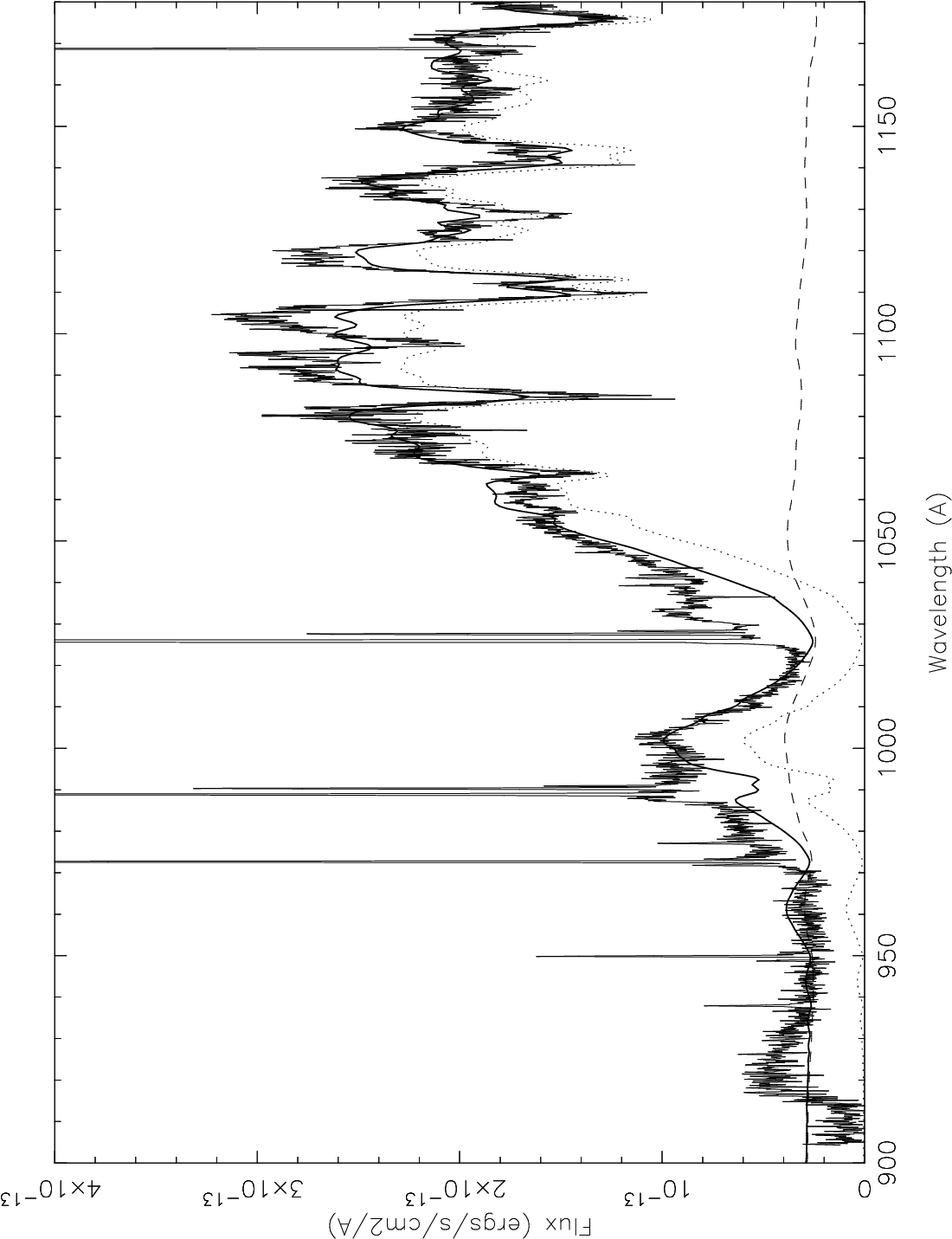}
\figurenum{5}
\end{figure}

\clearpage
\newpage 

\begin{figure}
\plotone{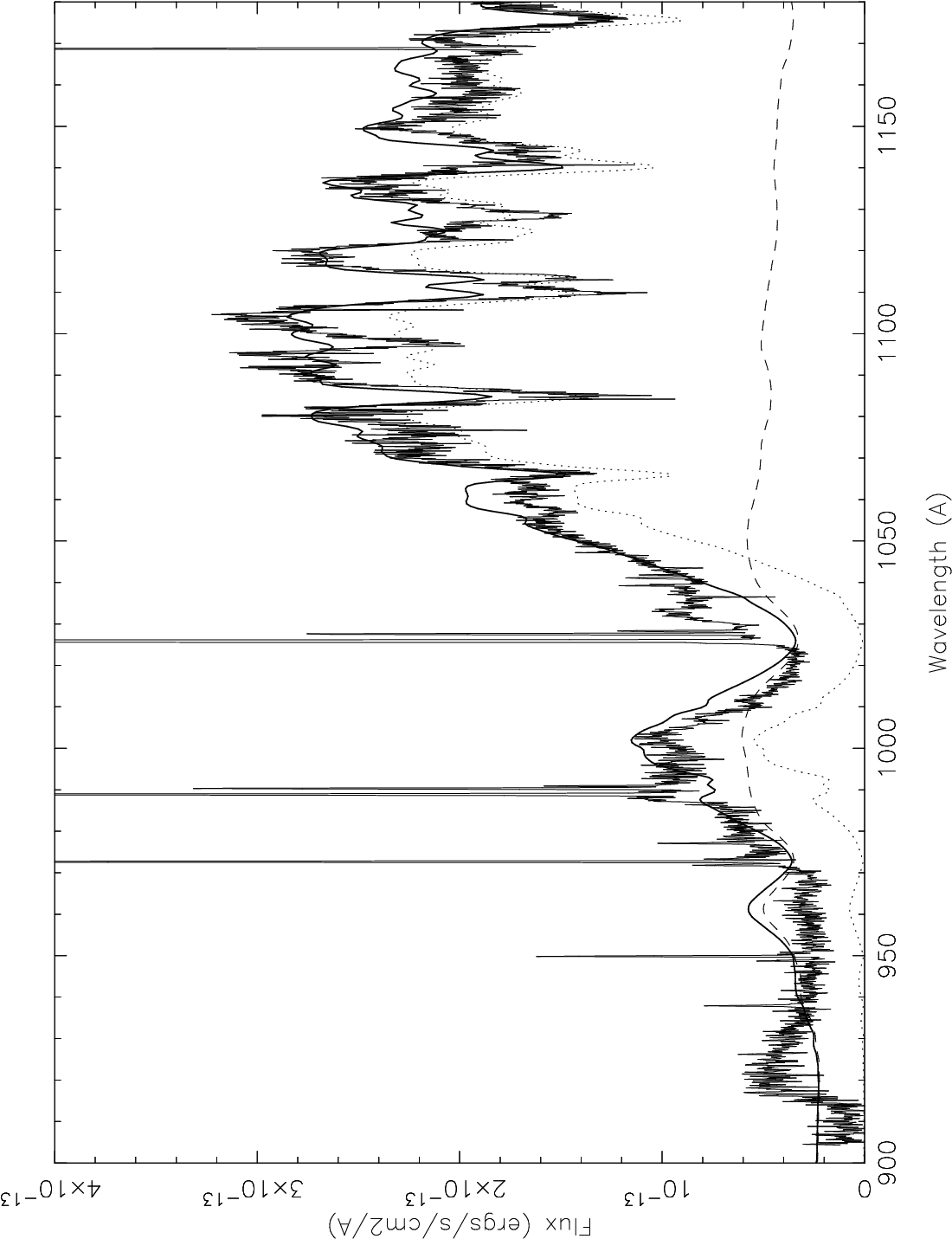}
\figurenum{6}
\end{figure}

\begin{deluxetable}{ccclc} 
\tablewidth{0pc}
\tablecaption{Line identification for the {\it{FUSE}} spectrum of VW Hyi} 
\tablehead{ 
$\lambda $ & EQW$^1$ & FWHM$^2$ & Line & Wavelength \\
  \AA\     &  \AA\   & \AA\     &      &    \AA    
} 
\startdata 
   1085.11  &     2.064 & 3.675 & \Ion{N}{II}   & 1085.7 \\
   1112.04  &     4.234 & 7.214 & \Ion{Si}{III} & 1113.2 \\
   1128.13  &     3.488 & 9.144 & \Ion{Fe}{III} & 1128.7 \\
            &           &       & \Ion{Fe}{II}  & 1129.2 \\
   1143.68  &     2.931 & 8.052 & \Ion{Si}{III} & 1144.3 \\
   1159.08  &     2.409 & 13.84 & \Ion{Si}{III} & 1160.2 \\
   1175.11  &     1.806 & 4.596 & \Ion{C}{III}  & 1174.9 \\
            &           &       &               & 1175.3 \\
            &           &       &               & 1175.6 \\
            &           &       &               & 1175.7 \\
            &           &       &               & 1176.0 \\
            &           &       &               & 1176.4
\enddata
\tablenotetext{1}
{In the second column: EQW is the line equivalent width in \AA.} 
\tablenotetext{2}
{In the third column: FWHM is the full width at half maximum in \AA.}
\end{deluxetable} 

\clearpage

\begin{deluxetable}{cccccccccccc} 
\tablewidth{0pc}
\tablecaption{Best fit models for VW Hyi} 
\tablehead{ 
model & n & $\chi^2_{\nu}$ & $T_{wd}$ & log g & $V_{rot}\sin{i}$ & Si & C & $M_{wd}$ & Log $\dot{M}$ & $i$ & $T_{belt}$  \\
 &    & & 1000K & cm/s$^2$ & km/s & $\odot$ & $\odot$ & $M_\odot$ & $M_{\odot}$yr$^{-1}$ & deg & 1000K  
} 
\startdata 
  WD    & 1  & 14.8 & 26 & 8.80 &  400  & 0.5  & 0.1 & 1.14 &       &     &      \\
  WD    & 2  & 32.0 & 22 & 8.37 &  400  & 1.0  & 1.0 & 0.86 &       &     &      \\
  WD    & 3  & 28.2 & 23 & 8.37 &  400  & 1.0  & 1.0 & 0.86 &       &     &      \\
  WD    & 4  & 81.5 & 24 & 8.37 &  400  & 1.0  & 1.0 & 0.86 &       &     &      \\
  WD    & 5  & 226.3& 25 & 8.37 &  400  & 1.0  & 1.0 & 0.86 &       &     &      \\
disk    & 6  & 19.9 &    &      &       &      &     & 1.03 & -10.5 & 60  &      \\ 
disk    & 7  &$10^4$&    &      &       &      &     & 0.80 & -9.5  & 60  &      \\
disk    & 8  & 29.8 &    &      &       &      &     & 0.80 & -10   & 60  &      \\
disk    & 9  & 286  &    &      &       &      &     & 0.80 & -10.5 & 60  &      \\
WD+disk & 10 & 11.0 & 23 & 8.99 &  400  & 0.5  & 0.1 & 1.21 & -10.5 & 81  &      \\
WD+disk & 11 & 32.9 & 21 & 8.23 &  400  & 1.0  & 1.0 & 0.80 & -10.5 & 60  &      \\
WD+belt & 12 & 7.06 & 23 & 8.54 &  400  & 2.0  & 0.2 & 0.96 &       &     & 48   \\
WD+belt & 13 & 10.9 & 22 & 8.37 &  400  & 1.0  & 1.0 & 0.86 &       &     & 50   \\
\enddata
\end{deluxetable} 

\clearpage

\begin{deluxetable}{lcccrl} 
\tablewidth{0pc}
\tablecaption{Some Recent Observations of VW Hyi} 
\tablehead{ 
Instrument  & $T_{eff}$ & $\dot{M}$ & Type$^1$ & Days$^2$ & Reference  \\
            &    1000K    &  $M_{\odot}/yr$  &     &       &  } 
\startdata
IUE      & 18.0 & $1 \times 10^{-10}$ &    &    Q & Wade et al. 1994 \\
HST/FOS  & 22.5 &                     & NO &  1-Q & Sion et al. 1996 \\
HST/FOS  &      & $3 \times 10^{-9}$  & SO &  5-O & Huang al. 1996 \\
HST/FOS  & 20.5 &                     & SO & 10-Q & Sion et al. 1996 \\
HUT      & 18.7 & $4 \times 10^{-11}$ & NO & 13-Q & Long et al. 1996 \\
HST/GHRS & 22.0 &                     & NO & 30-Q & Sion et al. 1997 \\
HST/STIS & 22.5 &                     & SO &  2-Q & Sion et al. 2001 \\
HST/STIS & 21.5 &                     & SO &  7-Q & Sion et al. 2001 \\
FUSE     & 23.0 &                     & NO & 11-Q & this work \\
\enddata

\tablenotetext{1}
{In the fourth column: NO - Normal Outburst; SO - Superoutburst.} 
\tablenotetext{2}
{In the fifth column: The number indicates how many days the 
observation was made into Q - quiescence or O - outburst.}
\end{deluxetable} 

\end{document}